\documentstyle[prl,preprint,aps]{revtex}
\input epsf

\begin{document}

\title{
$d_{x^2-y^2}$ Symmetry and the Pairing Mechanism
}
\author{ N.\ Bulut and D.J.\ Scalapino }
\address{
Department of Physics, University of California \\
Santa Barbara, CA 93106--9530
}
\date{Submitted in February 1995, in press, PRB}
\maketitle
\begin{abstract}
An important question is if the gap in the high temperature cuprates  
has $d_{x^2-y^2}$ symmetry, what does that tell us about the 
underlying interaction responsible for pairing.
Here we explore this by determining how three different types
of electron--phonon interactions affect the $d_{x^2-y^2}$ pairing
found within an RPA treatment of the 2D Hubbard model.
These results imply that interactions which become more positive 
as the momentum transfer increases favor $d_{x^2-y^2}$ pairing
in a nearly half--filled band.

\end{abstract}


\narrowtext

There have been a great deal of interest in the 
interpretation of recent experiments which address the question 
of the symmetry of the gap in the high temperature 
cuprate superconductors\cite{Levi,Schrieffer,Scalapino}.
If this symmetry turns out to be $d_{x^2-y^2}$,
it is natural to ask what this would imply about 
the pairing mechanism.
Here we discuss the relationship of the $d_{x^2-y^2}$ symmetry
to the pairing mechanism by examining how three types of 
electron--phonon interactions affect $d_{x^2-y^2}$ pairing. 

As discussed by a number of authors\cite{Levi,Schrieffer,Scalapino}, 
a $d_{x^2-y^2}$
gap naturally arises from the exchange of antiferromagnetic 
spin fluctuations.
However, the physical picture that emerges 
from these calculations
is more general and shows that a $d_{x^2-y^2}$ gap will occur
for a nearly half--filled band when there is an effective singlet 
interaction which is repulsive for on--site pairing and attractive 
for near--neighbor pairing.
This spatial structure of the interaction
means that its Fourier transform
becomes more positive as the 
momentum transfer increases towards large values.
This is easily understood from the BCS gap equation
\widetext
\begin{equation}
\Delta_{\bf p}= 
-\sum_{{\bf p}'}
{ V({\bf p}-{\bf p}') \Delta_{{\bf p}'} \over 
2E_{{\bf p}'} }.
\label{Dp}
\end{equation}
\narrowtext
Near half--filling of the 2D system, the phase space is such 
that 
the important scattering process take electrons from 
$({\bf p}\uparrow,{\bf -p}\downarrow)$
with ${\bf p}$ near a corner of the fermi surface, 
say near $(\pi,0)$,
to $({\bf p'}\uparrow,{\bf -p'}\downarrow)$
with ${\bf p'}$ near $(0,\pi)$ or $(0,-\pi)$. 
Since the interaction $V({\bf p}-{\bf p'})$ is positive,
the relative phase of the states
$({\bf p}\uparrow,{\bf -p}\downarrow)$
making up the bound Cooper pair changes sign as 
${\bf p}$  goes from $(\pi,0)$ to $(0,\pi)$ or $(0,-\pi)$
leading to a gap with $d_{x^2-y^2}$ symmetry.
This is the case within an RPA approximation
in which the interaction is mediated 
by the exchange of antiferromagnetic spin--fluctuations.
It has also been found by Monte Carlo 
calculations\cite{White} that for the Hubbard model
the pairing interaction is attractive in the $d_{x^2-y^2}$ channel.
In these cases the interaction is positive at all momentum transfers
becoming larger in the region near $(\pi,\pi)$ associated 
with the short range antiferromagnetic correlations.

In order to explore the effect of electron--phonon interactions, 
we begin with a 2D Hubbard model on a square lattice.
\widetext
\begin{equation}
H= 
- t \sum_{ \langle i,j\rangle, s} 
( c^{\dagger}_{is}c_{js} + 
 c^{\dagger}_{js}c_{is}  )
+ U\sum_{i} 
n_{i\uparrow}n_{i\downarrow },
\label{H}
\end{equation}
\narrowtext
Here $t$ is a near neighbor hopping and $U$ is the onsite 
Coulomb interaction. 
We take a simple phenomenological RPA form\cite{RPAform}
of the singlet pairing interaction 
associated with the exchange of spin fluctuations
\widetext
\begin{equation}
V_{\rm SF}(p'-p)= {3\over 2} {\overline U}^2\chi(p'-p)
\label{VSF}
\end{equation}
\narrowtext
with $\chi(q)=\chi_0(q)/[1-{\overline U}\chi_0(q)]$.
Here $p=({\bf p},i\omega_n)$, 
${\overline U}$ is a renormalized Coulomb interaction 
and $\chi_0$ is the spin susceptibility
\widetext
\begin{equation}
\chi_0({\bf q},\omega)= {1\over N}\sum_{\bf p} 
{ f(\varepsilon_{{\bf p}+{\bf q}})- f(\varepsilon_{\bf p})
\over 
\omega-(\varepsilon_{{\bf p}+{\bf q}} -\varepsilon_{\bf p} ) + i0^+ } 
\label{chi0}
\end{equation}
\narrowtext
with $\varepsilon_{\bf p}=-2t(\cos{p_x} + \cos{p_y})-\mu$.
Now, in addition to $V_{\rm SF}$, we will examine 
three model electron--phonon interactions.
The first is a simple on--site Holstein coupling of the form
\widetext
\begin{equation}
V_1 = \sum_i g x_i n_i
\label{Holstein}
\end{equation}
\narrowtext
with $x_i$ the atomic displacement
at site $\hat{i}$ and $n_{i}=n_{i\uparrow} + n_{i\downarrow}$
the onsite electron density. 
One could imagine this type of coupling arising from 
the interaction with an apical oxygen O(4).
In Eq.~(5) the coupling is linear in the atomic displacement
rather than quadratic.
This is possible since O(4) breaks the reflection symmetry with respect to
a single CuO$_2$ layer.
The second electron--phonon interaction can be viewed as arising 
from the in--plane breathing motion\cite{Song} of an O(2) oxygen
\widetext
\begin{equation}
V_2 = \sum_i g \big[ x_i (n_i-n_{i+x}) 
+ y_i (n_i-n_{i+y}) \big].
\label{breathing}
\end{equation} 
\narrowtext
Here $x_i$ describes the displacement of the O(2) along the 
${\bf x}$--axis between the Cu sites at $\hat{i}$ and 
$\hat{i}+{\bf \hat{x}}$ and 
$y_i$ the ${\bf y}$--axis displacement of an O(2) 
along the ${\bf y}$--axis 
between the Cu sites at $\hat{i}$ and $\hat{i}+{\bf \hat{y}}$.
The third interaction involves an axial ${\bf z}$--motion of a 
{\it buckled} O(2) atom.
\widetext
\begin{equation}
V_3 = \sum_i g \big[ z_i^x (n_i+n_{i+x}) + z_i^y (n_i+n_{i+y}) \big].
\label{buckling}
\end{equation}
\narrowtext
Here $z_i^x$ is for an O(2) between the $\hat{i}$ 
and $\hat{i}+{\bf \hat{x}}$ 
sites and $z_i^y$ is for an O(2) between the 
$\hat{i}$ and $\hat{i}+{\bf \hat{y}}$ sites.
Linear coupling of this type is possible for buckled Cu--O--Cu bonds.
The electron--phonon interactions considered here are diagonal
in the electron number representation.
We note that it would also be interesting to study the effects of 
phonon modes 
where the electron-phonon coupling is not diagonal
in the electron number representation.

Assuming for the discussion that the lattice coordinate can be
described as a local harmonic oscillator with frequency $\omega_0$
(which just as $g$ is of course different for the different modes), 
the effective electron--electron interaction
mediated by the exchange of these phonons is 
\widetext
\begin{equation}
V_{\rm ph}= 
{ -2|g({\bf q})|^2 \omega_0 \over 
\omega_m^2 + \omega_0^2 }
\label{Vph}
\end{equation}
\narrowtext
where  $\omega_m$ is the Matsubara frequency $2m\pi T$.
Here for the local interaction Eq.~(5)
\widetext
\begin{equation}
|g({\bf q})|^2 = { |g|^2 \over 2M\omega_0}
\label{glocal}
\end{equation}
\narrowtext
while for the breathing mode, Eq.~(6),
\widetext
\begin{equation}
|g({\bf q})|^2 = { 2|g|^2 \over 2M\omega_0}
( \sin^2{q_x/2} + \sin^2{q_y/2} )
\label{gbreathing}
\end{equation}
\narrowtext
and for the axial mode, Eq.~(7),
\widetext
\begin{equation}
|g({\bf q})|^2 = { 2|g|^2 \over 2M\omega_0}
( \cos^2{q_x/2} + \cos^2{q_y/2} )
\label{gbuckling}
\end{equation}
\narrowtext
with $M$ the O ion mass.

In order to see how these interactions affect the pairing we 
examine the leading eigenvalue $\lambda$ and eigenfunction 
$\phi(p)$ of the Bethe--Salpeter equation, 
neglecting self--energy contributions\cite{Self},
\widetext
\begin{equation}
\lambda \phi(p) = - {T\over N} \sum_{p'}
\big[ V_{\rm SF}(p-p') + V_{\rm ph}(p-p')\big]
G(p') G(-p') \phi(p'),
\label{BS}
\end{equation}
\narrowtext
where $G(p)$ is the single--particle Green's function given by
\widetext
\begin{equation}
G(p)= {1 \over
i\omega_n - \varepsilon_{p}}.
\label{Gp}
\end{equation}
\narrowtext
Here and in the following we will measure energies in units of $t$.
The chemical potential has been chosen so that the site occupation 
$\langle n_{i\uparrow} + n_{i\downarrow}\rangle =0.875$
and an effective Coulomb interaction ${\overline U}=2$
has been taken.
We will also take $\omega_0=0.25$ and 
$|g|^2/M\omega_0 = 1$
corresponding to 
an electron--phonon coupling strength 
$|g|^2N(0)/M\omega_0^2 \simeq 0.4$,
where $N(0)$ is the electron density of states at $\mu_F$.

We find that the leading eigenvalue in the even frequency 
singlet channel has $d_{x^2-y^2}$ symmetry and the temperature 
dependence of the eigenvalue $\lambda_{x^2-y^2}(T)$
is shown in Figures 1 and 2 for the 
electron--phonon interactions
given by Eqs.~(6) and (7), respectively.
The solid line in each figure shows the eigenvalue in the 
absence of the phonon mediated interaction ($g=0$), while the 
dashed curve shows the effect of including the phonon mediated term.
It is clear that the breathing mode interaction, 
Eqs.(6) and (10), suppress 
$d_{x^2-y^2}$ pairing\cite{Schuttler}
while the axial O(2) mode of Eqs.(7) and (11) enhances
the $d_{x^2-y^2}$
pairing, raising $T_c$.
The local interaction,
Eqs.~(5) and (9), is orthogonal to the $d_{x^2-y^2}$ gap and hence
does not affect the $d_{x^2-y^2}$ eigenvalue when self--energy 
effects are neglected.
Including it in the self--energy will act to suppress $T_c$ due to
the wave function renormalization\cite{Self}.
To understand the behavior shown in Figs. 1 and 2 we
note that the strength of the 
coupling to the axial mode, Eq.(11), decreases as ${\bf q}$
approaches $(\pi,\pi)$.
Because the phonon mediated interaction, Eq.(7), is negative, 
decreasing the magnitude of the coupling 
$|g({\bf q})|^2$ acts to make the interaction 
{\it more positive}   as the momentum transfer increases.
As discussed in the introduction, this is the criteria for a
$d_{x^2-y^2}$ gap to form when the system is near half--filling. 
Clearly it will be interesting to examine the isotope effect 
within models in which the strength of the electron--phonon coupling
decreases at large momentum transfers.

Thus we conclude that a $d_{x^2-y^2}$ gap implies that the 
pairing interaction becomes more positive for large momentum transfers.
This is clearly the case for the spin--fluctuation interaction,
Eq.(2), but as shown it can also occur for the attractive phonon mediated 
interaction if $|g({\bf q})|^2$ {\it decreases} at large 
momentum transfers.
This form of coupling would also give rise to an electron--phonon
coupling constant $\lambda$ which could be large 
compared to the effective coupling constant $\lambda_{\rm tr}$
entering transport processes
since the transport coupling constant $\lambda_{\rm tr}$
weights large momenta transfers more heavily\cite{Crespi,Zeyher}.
For small momentum transfers ${\bf q}$,
a scattering of $({\bf p}\uparrow,-{\bf p}\downarrow)$
to $({\bf p}+{\bf q}\uparrow,-{\bf p}-{\bf q}\downarrow)$
with ${\bf p}$ near a corner of the Fermi surface connects 
regions which have the same sign of the $d_{x^2-y^2}$ gap
so that according to Eq.~(1) an attractive 
electron--phonon interaction (negative $V(q)$) enhances 
$\Delta_{\bf p}$.
Another way to see that this latter case is similar to
the spin--fluctuation interaction is to add $U$ onto the phonon 
interaction giving $U+V_{\rm ph}$.
For the $d_{x^2-y^2}$ channel, a constant has no effect
but if it is 
larger than the magnitude of the phonon mediated interaction, 
Eq.(7), then as $|g({\bf q})|^2$ decreases, the total interaction 
$U+V_{\rm ph}$ is positive 
and increases as ${\bf q}$ becomes large,
just as $V_{\rm SF}$.
In order to obtain more quantitative information on the 
role of the electron--phonon interaction, it would be useful 
to have band structure calculations\cite{Bandstructure}
of the $d_{x^2-y^2}$electron--phonon coupling constant for the 
$\nu$--mode
\widetext
\begin{equation}
\lambda_{d_{x^2-y^2}}^{\nu} = 
{ 2 \sum_{k,k'} g(k) g(k')
{ |M^{\nu}_{kk'}|^2 \over \omega_{k-k'} }
\delta(\varepsilon_k) \delta(\varepsilon_{k'})
\over 
\sum_{\bf k} g^2(k) \delta(\varepsilon_k) }.
\label{lambda}
\end{equation}
\narrowtext
Here $g(k)\sim (\cos{k_x} - \cos{k_y})$, 
$\varepsilon_{k}$ is the band energy with ${k}$ including 
the band index and $M_{kk'}^{\nu}$ is the electron--phonon 
matrix element for the $\nu$'th phonon mode\cite{Mkk}.

\vskip 0.2in
\centerline{{\bf Acknowledgments}}
The authors gratefully acknowledge support from the National 
Science Foundation under Grant No. DMR92--25027.
The numerical computations reported in this paper were performed at the 
San Diego Supercomputer Center.


\begin{figure}
\centerline{\epsfysize=8cm \epsffile[18 184 592 598] {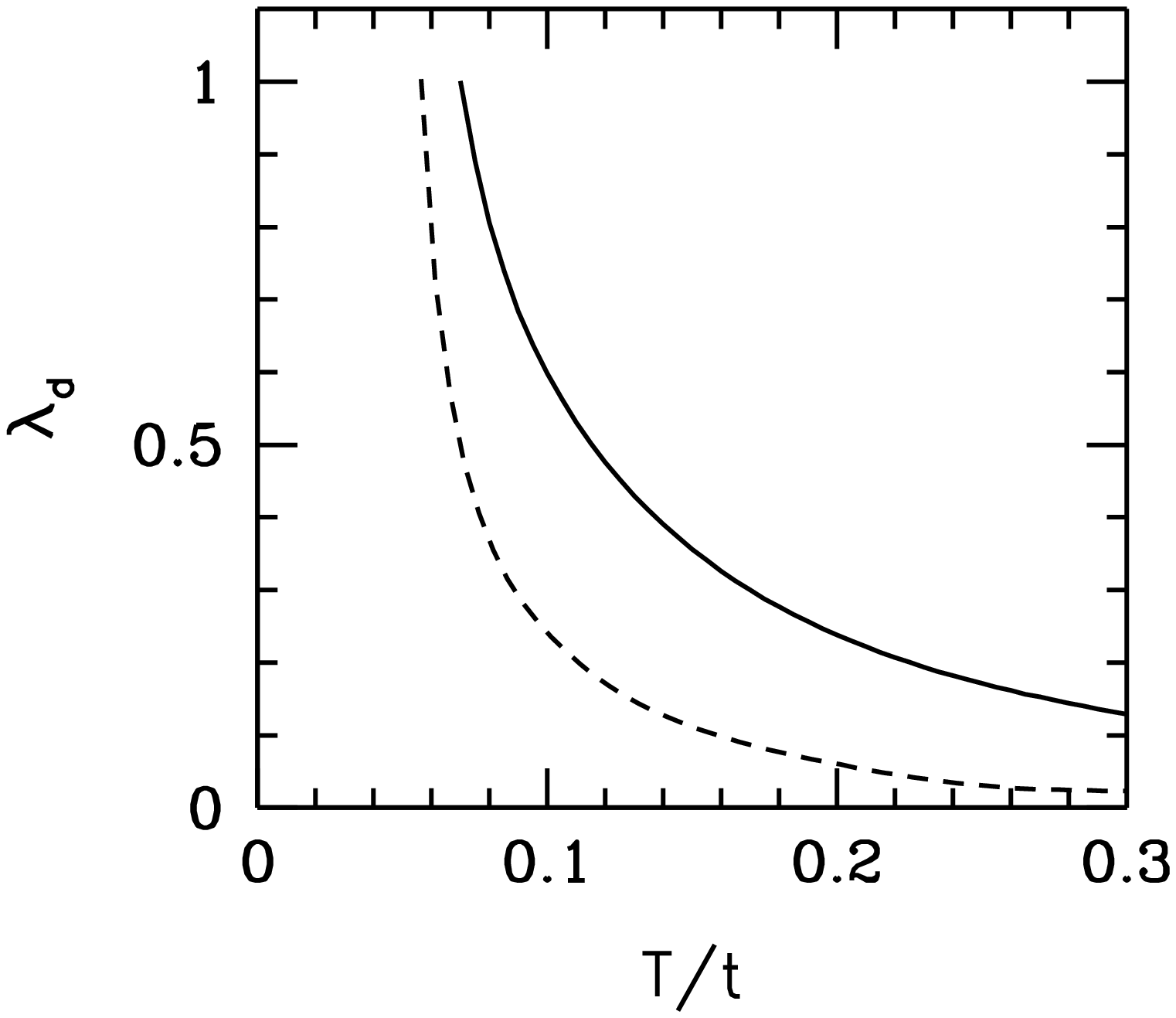}}
\vspace{0.5cm}
\caption{
The $d_{x^2-y^2}$ eigenvalue of Eq.~(2) versus the temperature $T$ 
in units of the hopping $t$.
The solid curve gives the eigenvalue for just the 
spin--fluctuation interaction $V_{\rm SF}$ 
and the dashed curve shows the effect when the electron--phonon
interaction $V_2$ associated with the breathing mode,
Eqs.~(6) and (10), is added to $V_{\rm SF}$.
The coupling constants are given in the text.
}
\label{fig1}
\end{figure}
  
\begin{figure}
\centerline{\epsfysize=8cm \epsffile[18 184 592 598] {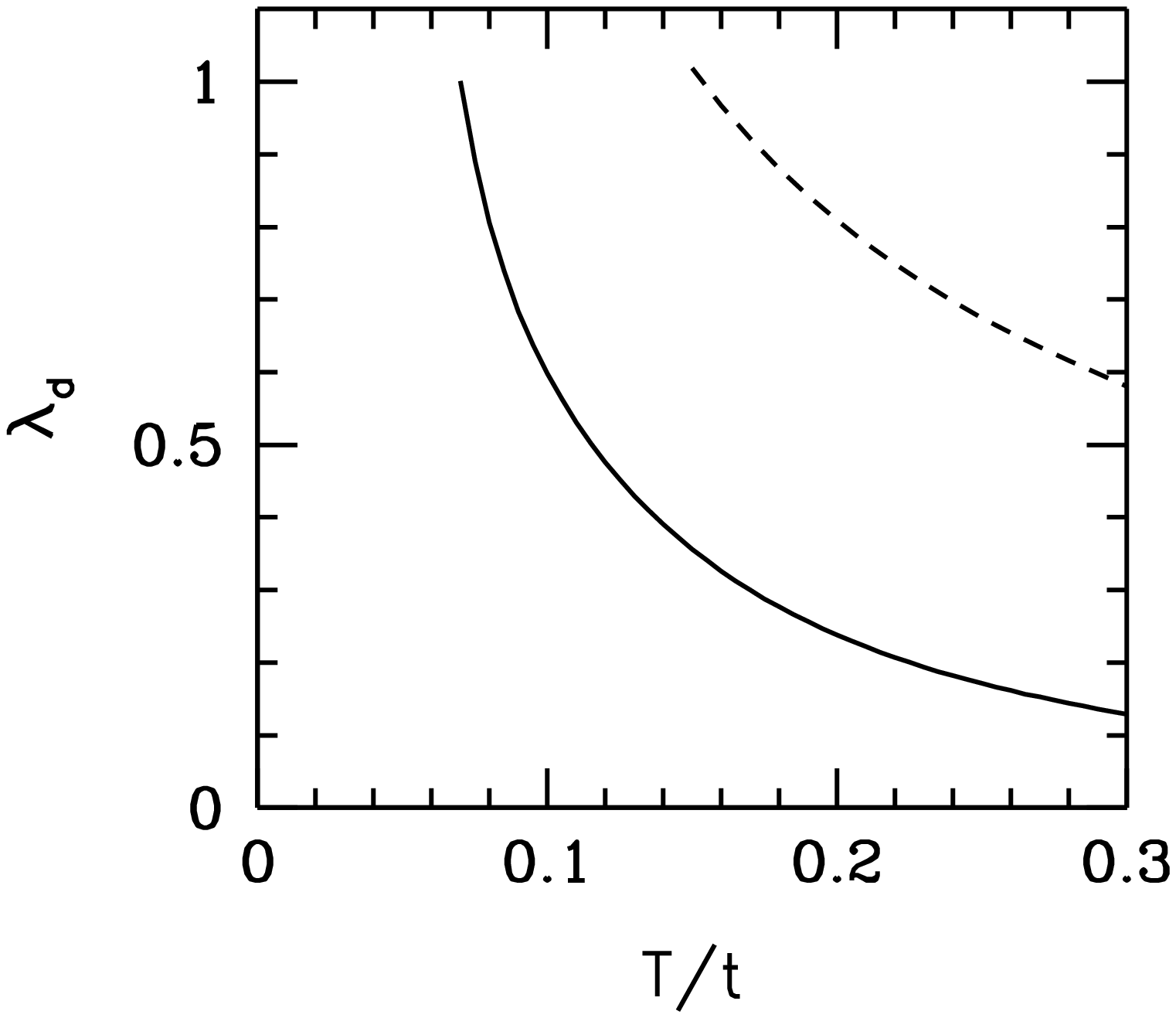}}
\vspace{0.5cm}
\caption{
Same as Fig. 1 except for the axial O electron--phonon 
interaction $V_3$, Eqs.~(7) and (11).
}
\label{fig2}
\end{figure}


\begin{references}
\bibitem{Self} The effect of the self--energy is to reduce 
the effective strength of the interactions due to wave 
function renormalization.  
Here we are interested in obtaining a qualitative 
understanding of the effect of the different 
electron--phonon interactions on the $d_{x^2-y^2}$ pairing 
channel and have neglected the self--energy and 
treated the coupling strength $g$ as a parameter.
\bibitem{Mkk} For the cuprates it will also be important to 
take into account the strong electron--electron 
correlations in computing $M_{kk'}^{\nu}$.
\end{references}
\end{document}